\documentclass[12pt,aps,prb,preprint]{revtex4-1}   

\usepackage{amsmath}    
\usepackage{graphicx}   
\usepackage{hyperref}
\newcommand{\sch}{Schr\"{o}dinger equation}
\begin{document}

\title{Directly integrating the \sch\ to determine tunneling rates for arbitrary one-dimensional potential barriers}
\author{Andy Rundquist}
 \affiliation{Hamline University, Department of Physics, St.\ Paul,
MN 55104}
 \email{arundquist@hamline.edu}   

\date{\today}

\begin{abstract}
By directly integrating the \sch\ starting in the transmission region and working backwards through the barrier, the tunneling probability can be determined for arbitrary potential barriers.  The method employs techniques familiar to undergraduates and is used here to study resonant tunneling.
\end{abstract}

\maketitle

\section{Introduction}

Quantum tunneling is a favorite conceptual topic for students.  It is a notion of something so very different from what is expected classically that can be described so easily by invoking memories of throwing balls at walls. Students are encouraged to find connections with frustrated total internal reflection in order to further cement their understanding of matter as waves.  Both in optics and quantum mechanics instructors can note how the continuity of the wave function and its derivative across boundaries implies that the wave cannot abruptly go to zero.  This enables students to see why a (typically small) portion of the wave can tunnel through a barrier.

On the other hand, the quantitative aspects of tunneling are a different story for students.  As usual, very simple situations can be calculated by hand like the square barrier but more typical barriers found in the lab require the use of a computer and an algorithm that can apply the conceptual physics they have learned (the boundary conditions) in an iterative fashion.

In this work present a different algorithm to calculate the tunneling probability of a particle with known energy through an arbitrary one-dimensional potential barrier.  It is both fast and accurate but it also uses tools that most students are familiar with in their studies of the \sch.  Specifically it involves the direct integration of the \sch\ in a manner very similar to the shooting method employed to find the eigenstates of an arbitrary potential well.  However, instead of needing to adjust parameters to find a particular eigenstate, students can directly inspect the results for any given particle energy and determine both the tunneling probability and the shape and nature of the wavefunction inside the barrier.

\section{Other methods}
The most common approach to calculating tunneling probabilities is to consider the barrier to be a collection of square barriers.  In the WKB approach, only the exponentially decaying portion of the wavefunction is kept and integrated through all the slices.\cite{simmons1963electric}  In the matrix transfer method, the boundary conditions among all the slices are carefully calculated.\cite{alexopoulos2007quantum,mendez:143,morelhao2007,probst2002tunneling,zhang2000tunnelling}  Specifically, at every boundary between the square slices the wavefunction and its slope are continuous.  In each slice the wavefunction is composed of two components: either a right and left traveling wave with a wavelength determined from the kinetic energy (the difference between the total energy and the barrier height); or a growing and decaying exponential whose growth rate is determined from the (negative) kinetic energy. Often these boundary condition equations are described in a matrix formalism as they are simple linear equations relating the incoming and outgoing wavefunctions along with the barrier heights of the slices.  The effect on the incoming wave by the barrier is then modeled by a single matrix that can used to solve for the tunneling probability.

There are also some approaches in the literature that have more directly integrated the \sch\ but all do a forward propagation as opposed to the backward one described below.\cite{ban2000novel,yunpeng1996new}  These approaches use both numeric and analytical methods to determine the phase of the incoming wave that enables solely a right-traveling wave in the transmission region.  The method below does not require such adjustments and simply gives both the wavefunction in the tunneling region and the tunneling probability after a single direct integration.

\section{Method}

Consider a tunneling situation as laid out in Figure~\ref{fig:arb}.  The first and third regions have a constant potential while the middle region can have any form, including discontinuities and regions where the particle is classically allowed.   Region I can have right- and left-traveling waves
\begin{equation}
\label{eqn: region I}
\psi_\text{I}=Ae^{i k_\text{I} x}+Be^{-i k_\text{I} x}
\end{equation}
while Region III only has a right traveling wave
\begin{equation}
\label{eqn: region III}
\psi_\text{III}=Fe^{i k_\text{III} x}
\end{equation}
where
\begin{equation}
\label{eqn: kI}
k_\text{I}=\sqrt{(2m^2/\hbar^2)E}
\end{equation}
and
\begin{equation}
\label{eqn: kIII}
k_\text{III}=\sqrt{(2m^2/\hbar^2)\left(E-V_\text{III}\right)}.
\end{equation}

Using a fourth-order Runge-Kutta technique we numerically integrate the real and imaginary parts of the \sch\ from the right edge of Region II ($x=L$) to the left edge ($x=0$).  Note that since the \sch\ does not have any single derivatives in it a Numeroff approach can also be used.\cite{noumerov1924method}  Note also that in \emph{Mathematica} you can integrate complex numbers with just one call to the Runge-Kutta solver (NDSolve). Since both the wavefunction and its slope will be the same on both sides of the boundary between Regions~II and III, the initial conditions are determined by arbitrarily setting $F=1$ and using the form from Region III: 
\begin{equation}
\label{eqn: init value}
\psi(L)=e^{i k_{\text{III}}L}\quad \text{and}\quad \psi'(L)=i k_{\text{III}}e^{i k_{\text{III}}L}.
\end{equation}

To determine the transmission probability, $T$, we need to find the value of $A$:
\begin{equation}
\label{eqn: T}
T=\frac{k_{\text{III}}}{k_{\text{I}}}\left|\frac{F}{A}\right|^{2}=\frac{k_{\text{III}}}{k_{\text{I}}}\frac{1}{\left|A\right|^{2}}.
\end{equation}
This is done by investigating the value of the wavefunction and its slope at $x=0$ where, according to Eq.~\ref{eqn: region I}, 
\begin{equation}
\label{eqn: psi_at_0}
\psi(0)=A+B\quad \text{and}\quad \psi'(0)=ik_{\text{I}}(A-B).
\end{equation}
Once again we have used the equality of the value and slope of the wavefunction across a boundary.

Combining Eqs.~\ref{eqn: T} and \ref{eqn: psi_at_0} yields
\begin{equation}
\label{eqn: full T}
T=\frac{k_{\text{III}}}{k_{\text{I}}} \frac{4}{\left| \psi(0)-i\psi'(0)/k_{\text{I}}\right|^{2}}.
\end{equation}
Once the \sch\ has been numerically integrated, the transmission probability is easily calculated.

This method employs many techniques used when teaching the numerical solution of eigenstates for arbitrary energy wells.  In those situations students are taught to employ the shooting method to find energies that produce physically allowable wavefunctions.  Students start with an initial value and slope (these can usually be pretty arbitrary), typically on one side of the energy well to be considered, and use a numeric integrator to predict the shape of the wavefunction in the well.  They then adjust the energy involved, which changes the shape, and they reject any energies that cause the wavefunction to asymptotically approach infinity at the other side of the well.  In this new application the initial value and slope are set by Eq.~\ref{eqn: init value} and the solution at every energy is used to calculate the transmission probability using Eq.~\ref{eqn: full T}. The two major differences in this new application are that the integration is done backwards spatially and that both the real and imaginary parts need to be integrated, as is illustrated in Figure~\ref{fig:realandimag}.  If you only do the real part (as is often done in the shooting method application) you are unable to calculate the transmission coefficient as seen in Eq.~\ref{eqn: full T}.

\section{Examples and comparisons}
\begin{figure}[h]
\includegraphics{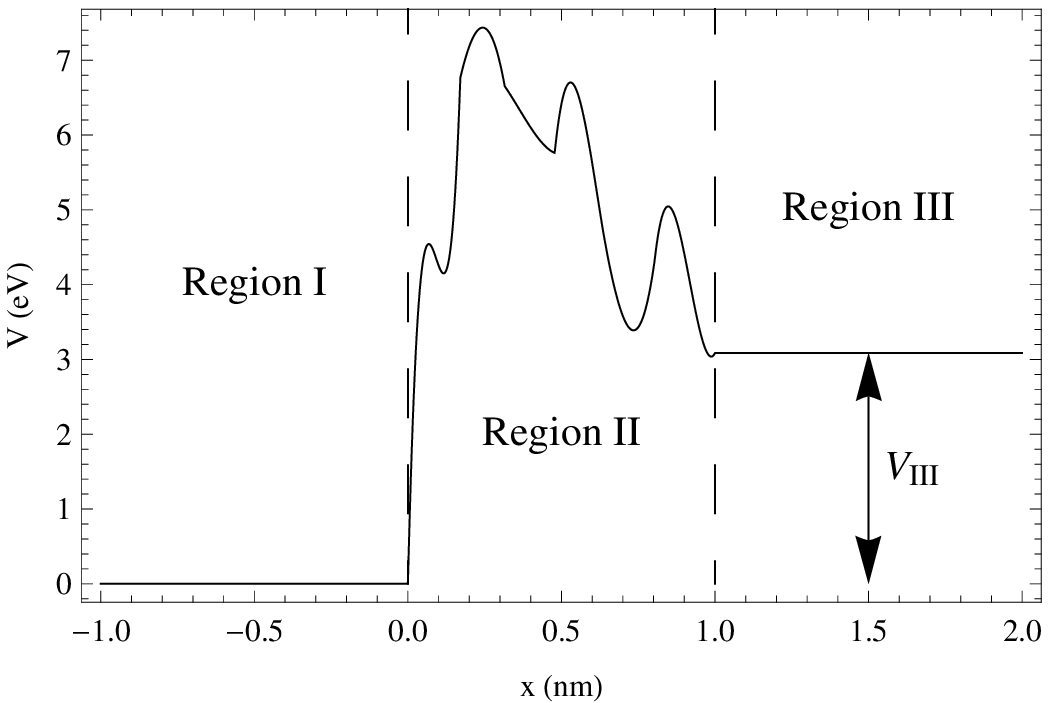}
\caption{\label{fig:arb}Arbitrary potential barrier.  The electron approaches in Region I and tunnels through Region II into Region III.}
\end{figure}

\begin{figure}[htbp]
   \includegraphics{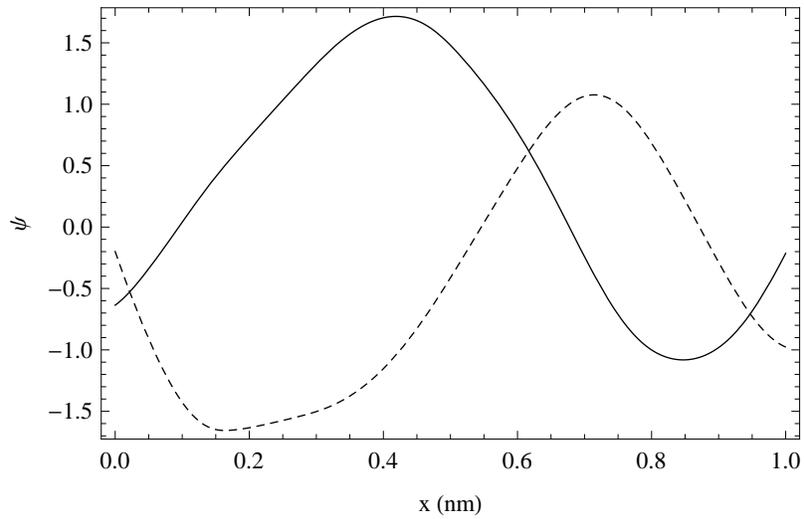} 
   \caption{\label{fig:realandimag}The real (solid) and imaginary (dashed) parts of the tunneling wavefunction for the barrier shown in Figure~\ref{fig:arb} for a particle energy of 7.5~eV.  This yields a tunneling probability of 93\%.}
\end{figure}

\begin{figure}[htbp]
   \includegraphics{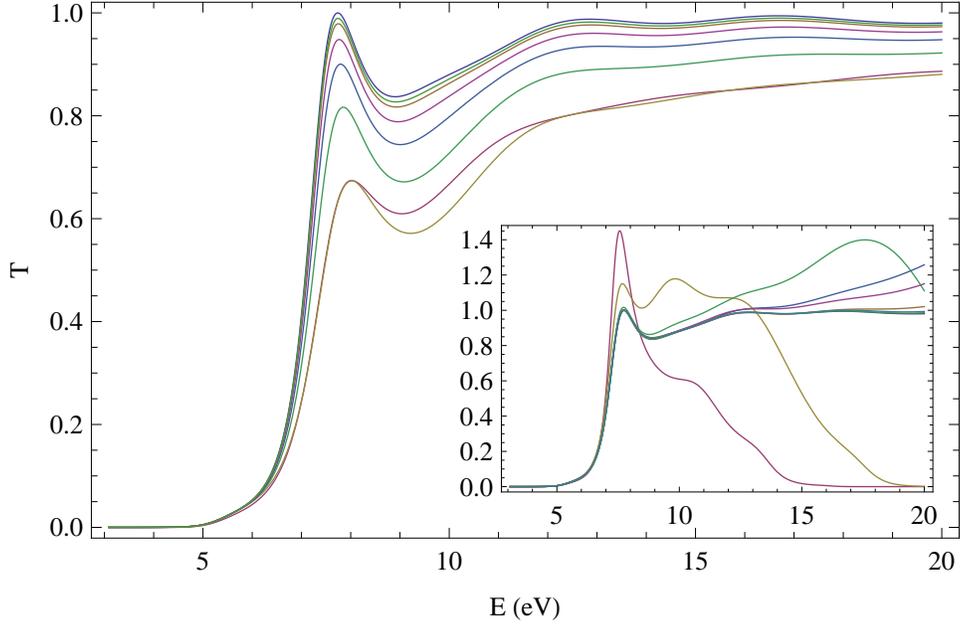} 
   \caption{\label{fig:compareinset}The transmission probability, $T$, as a function of particle energy for the potential barrier shown in Figure~\ref{fig:arb}.  This compares the current method with the matrix method with the number of slices set to 10, 20, 50, 100, 200, 500, 1000.  The current method is the top curve.  In the inset the current method is limited to a total number of steps ranging from five through twelve.}
\end{figure}

The transmission coefficient ($T$) as a function of particle energy for the potential shown in Figure~\ref{fig:arb} is given in Figure~\ref{fig:compareinset}.  The top curve is the result of the current method while the lower lines use the transfer matrix method with varying number of slices of Region~II.  Ultimately both approaches converge to the same result at every energy.

It is interesting to compare the transfer matrix method with the new method where the number of slices is compared with the number of steps that the Runge-Kutta method employs.  The transmission probability versus energy for the arbitrary barrier shown in Figure~\ref{fig:arb} is given for the total step number ranging from 5 to 12 in the inset of Figure~\ref{fig:compareinset}.  The curve with 300 steps is also shown.  It is clear that the number of steps needed for the Runge-Kutta method is far less than the number of slices needed in the transfer matrix method to achieve the same accuracy.  Note, however, that one should really compare the number of calculations involved when doing these comparisons.  A fairer comparison would need to multiply the number of Runge-Kutta steps by four, though this still shows that the current approach compares favorably to the tranfer matrix method.

\section{Resonant tunneling}
As an example to show the pedagogical uses of the current method, we consider resonant tunneling.  Specifically we compare the wavefunction in the barrier region to the eigenstates expected for a simply-shaped barrier.

\begin{figure}[htbp]
   \includegraphics{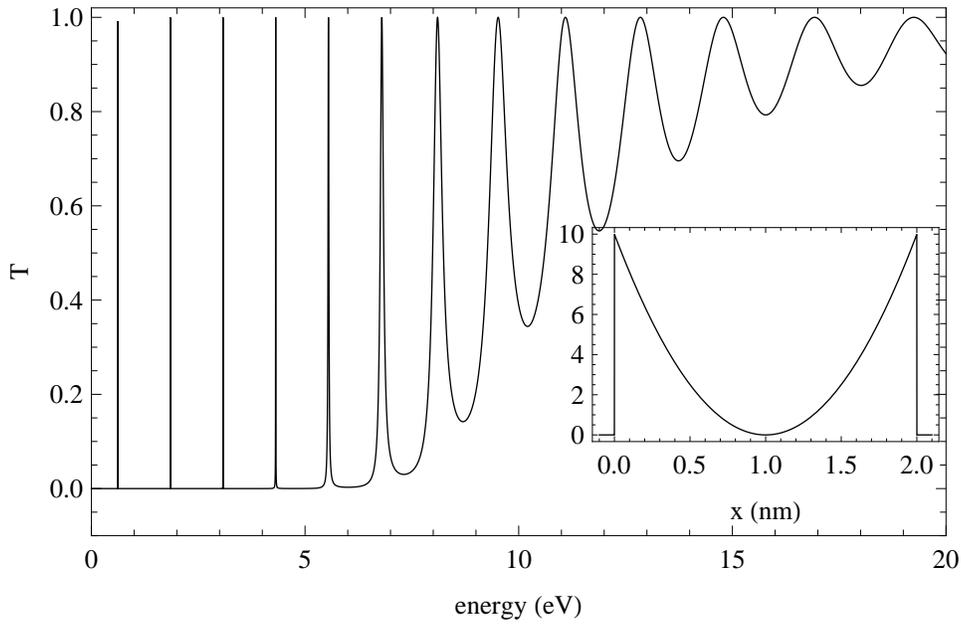} 
   \caption{\label{fig:resinset}The tunneling probability versus energy for the parabolic potential shown in the inset.}
\end{figure}

Consider the potential shown in the inset of Figure~\ref{fig:resinset}.  This parabolic potential barrier is given in electron volts by
\begin{equation}
V(x)=\left\{\begin{array}{ll}
	10 (x-1)^2	&	\text{if }0<x<2 \\
	0						&	\text{otherwise}
\end{array}
\right..
\label{eq:parabpotential}
\end{equation}
The analogous potential well that is not truncated has resonant energies at 
\begin{equation}
E_n=\left(n+\frac{1}{2}\right)\hbar \omega = \left(n+\frac{1}{2}\right) 1.232 \text{ eV}\quad \text{where }n=0,1,2,\ldots.
\label{eq:parabenergies}
\end{equation}
The transmission probability as a function of energy is shown in Figure~\ref{fig:resinset}.  The resonance peaks shown correspond very nearly with the eigenenergies of a parabolic well.  At lower energies where the resonance peaks are very sharp the energies are the same as the eigenenergies.  As the peaks become broader, the resonant energies become larger than the eigenenergies by as much as 20\% for $n=11$.  

\begin{figure}[htbp]
   \includegraphics{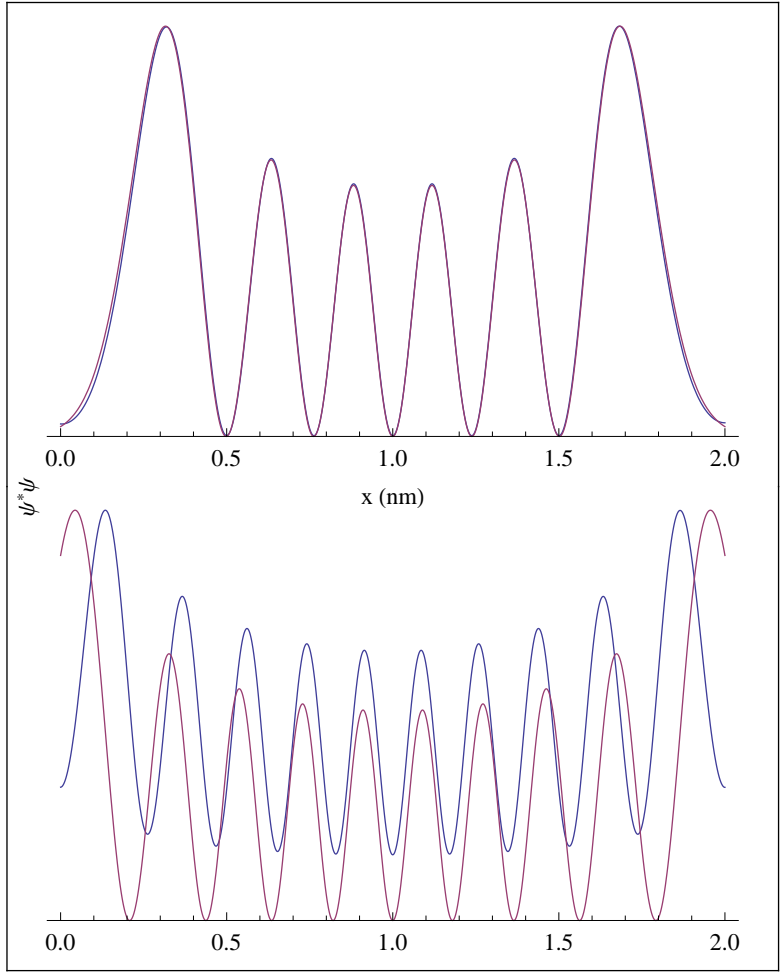} 
   \caption{\label{fig:parabcompare}Comparison of the resonant tunneling wavefunction with the eigenfunction for a parabola for $n=5$~(top) and $n=9$~(bottom).  The probability density ($\psi^*\psi$) is plotted as a function of position inside the potential barrier (shown in Figure~\ref{fig:resinset}).  The curves are all normalized to a max of unity.  The resonant wavefunction has higher energy so that the curve can bend to better meet the boundary conditions.}
\end{figure}

The reason the resonant energies grow larger than the eigenenergies as the energy increases is due to the boundary conditions that the wavefunction has to match at $x=0$ and $x=2$.  This can be seen by comparing the resulting wavefunctions (both the tunneling wavefunction and the eigenfunction for the parabola) as seen in Figure~\ref{fig:parabcompare}.  Close inspection of the wavefunctions near the boundaries indicates the differences between the tunneling wavefunction and the eigenstate.  While the eigenstate is decaying to zero in all cases, the tunneling wavefunction is forced to match the boundary condition at the right edge.  At low energies there is little difference as the exponential rise is very steep but at higher energies there is a higher bend needed that explains the rise in the energy compared to the analogous eigenenergy.

\section*{Conclusion}
We have discussed a new method for calculating both transmission probabilities and wavefunctions for a particle tunneling through an arbitrary one-dimensional barrier.  The approach is applicable at the undergraduate level as it uses common tools related to the shooting method for finding potential well eigenstates.  It is fast and accurate and enables the study of complex phenomena like resonant tunneling.

\bibliography{tunneling}

\newpage
\section*{Figures}

\end{document}